\documentclass[runningheads]{llncs}
\usepackage{graphicx}
\usepackage{comment}
\usepackage{amsmath,amssymb} 
\usepackage{color}
\usepackage{multirow}

\graphicspath{{./figures/}}
\DeclareGraphicsExtensions{.pdf,.jpeg,.png,.jpg,.eps}

\begin{document}
\pagestyle{headings}
\mainmatter
\def\ECCVSubNumber{7202}  

\title{Spatial-Adaptive Network for Single Image Denoising} 

\titlerunning{Spatial-Adaptive Network for Single Image Denoising}
%
\author{Meng Chang\inst{1} \and
Qi Li\inst{1} \and
Huajun Feng\inst{1} \and
Zhihai Xu\inst{1}}
\authorrunning{M. Chang et al.}
%
\institute{State Key Lab for Modern Optical Instruments, Zhejiang University, China\\
\email{\{changm, liqi, fenghj, xuzh\}@zju.edu.cn}}
\maketitle

\begin{abstract}
   Previous works have shown that convolutional neural networks can achieve good performance in image denoising tasks. However, limited by the local rigid convolutional operation, these methods lead to oversmoothing artifacts. A deeper network structure could alleviate these problems, but at the cost of additional computational overhead. In this paper, we propose a novel spatial-adaptive denoising network (SADNet) for efficient single image blind noise removal. To adapt to changes in spatial textures and edges, we design a residual spatial-adaptive block. Deformable convolution is introduced to sample the spatially related features for weighting. An encoder-decoder structure with a context block is introduced to capture multiscale information. By conducting noise removal from coarse to fine, a high-quality noise-free image is obtained. We apply our method to both synthetic and real noisy image datasets. The experimental results demonstrate that our method outperforms the state-of-the-art denoising methods both quantitatively and visually.
\keywords{image denoising, image restoration, image processing}
\end{abstract}

\section{Introduction}

Image denoising is an important task in computer vision. During image acquisition, noise is often unavoidable due to imaging environment and equipment limitations. Therefore, noise removal is an essential step, not only for visual quality but also for other computer vision tasks. Image denoising has a long history, and many methods have been proposed. Many of the early model-based methods found natural image priors and then applied optimization algorithms to solve the model iteratively~\cite{osher2005iterative,aharon2006k,xu2007iterative,zoran2011learning}. However, these methods are time consuming and cannot effectively remove noise. With the rise of deep learning, convolutional neural networks (CNNs) have been applied to image denoising tasks and have achieved high-quality results. 


On the other hand, the early works assumed that noise is independent and identically distributed. Additive white Gaussian noise (AWGN) is often adopted to create synthetic noisy images. People now realize that noise presents in more complicated forms that are spatially variant and channel dependent. Therefore, some recent works have made progress in real image denoising~\cite{Ploetz2018NNN,zhou2019awgn,guo2019toward,anwar2019real}. 

However, despite numerous advances in image denoising, some issues remain to be resolved. A traditional CNN can use only the features in local fixed-location neighborhoods, but these may be irrelevant or even exclusive to the current location. Due to their inability to adapt to textures and edges, CNN-based methods result in oversmoothing artifacts and some details are lost. In addition, the receptive field of a traditional CNN is relatively small. Many methods deepen the network structure~\cite{tai2017memnet} or use a non-local module to expand the receptive field~\cite{liu2018non,zhang2019residual}. However, these methods lead to high computational memory and time consumption, hence they cannot be applied in practice. 

In this paper, we propose a spatial-adaptive denoising network (SADNet) to address the above issues. A residual spatial-adaptive block (RSAB) is designed to adapt to changes in spatial textures and edges. We introduce the modulated deformable convolution in each RSAB to sample the spatially relevant features for weighting. Moreover, we incorporate the RSAB and residual blocks (ResBlock) in an encoder-decoder structure to remove noise from coarse to fine. To further enlarge the receptive field and capture multiscale information, a context block is applied to the coarsest scale. Compared to the state-of-the-art methods, our method can achieve good performance while maintaining a relatively small computational overhead. 

In conclusion, the main contributions of our method are as follows: 
\begin{itemize}
  \item We propose a novel spatial-adaptive denoising network for efficient noise removal. The network can capture the relevant features from complex image content, and recover details and textures from heavy noise.  
  \item We propose the residual spatial-adaptive block, which introduces deformable convolution to adapt to spatial textures and edges. In addition, using an encoder-deocder structure with a context block to capture multiscale information, we can estimate offsets and remove noise from coarse to fine.
  \item We conduct experiments on multiple synthetic image datasets and real noisy datasets. The results demonstrate that our model achieves state-of-the-art performances on both synthetic and real noisy images with a relatively small computational overhead.
\end{itemize}

\section{Related works}

In general, image denoising methods include model-based and learning-based methods. Model-based methods attempt to model the distribution of natural images or noise. Then, using the modeled distribution as the prior, they attempt to obtain clear images with optimization algorithms. The common priors include local smoothing~\cite{osher2005iterative,xu2007iterative}, sparsity~\cite{aharon2006k,mairal2009non,xu2018trilateral}, non-local self-similarity~\cite{buades2005non,dabov2007image,dabov2007color,xu2017multi,gu2014weighted} and external statistical prior~\cite{zoran2011learning,xu2018external}. Non-local self-similarity is the notable prior in the image denoising task. This prior assumes that the image information is redundant and that similar structures exist within a single image. Then, self-similar patches are found in the image to remove noise. Many methods have been proposed based on the non-local self-similarity prior including NLM~\cite{buades2005non}, BM3D~\cite{dabov2007image,dabov2007color}, and  WNNM~\cite{gu2014weighted,xu2017multi}, all of which are currently widely used.

With the popularity of deep neural networks, learning-based denoising methods have developed rapidly. Some works combine natural priors with deep neural networks. TRND~\cite{chen2016trainable} introduced the field-of-experts prior into a deep neural network. NLNet~\cite{lefkimmiatis2017non} combined the non-local self-similarity prior with a CNN. Limited by the designed priors, their performance is often inferior compared to end-to-end CNN methods. DnCNN~\cite{zhang2017beyond} introduced residual learning and batch normalization to implement end-to-end denoising. FFDNet~\cite{zhang2018ffdnet} introduced the noise level map as the input and enhanced the flexibility of the network for non-uniform noise. MemNet~\cite{tai2017memnet} proposed a very deep end-to-end persistent memory network for image restoration, which fuses both short-term and long-term memories to capture different levels of information. Inspired by the non-local self-similarity prior, a non-local module~\cite{wang2018non} was designed for neural networks. NLRN~\cite{liu2018non} attempted to incorporate non-local modules into a recurrent neural network (RNN) for image restoration. N3Net~\cite{Ploetz2018NNN} proposed neural nearest neighbors block to achieve non-local operation. RNAN~\cite{zhang2019residual} designed non-local attention blocks to capture global information and pay more attention to the challenging parts. However, non-local operations lead to high memory usage and time consumption. 

Recently, the focus of researchers has shifted from AWGN to more realistic noise. Some recent works have made progress on real noisy images. Several real noisy datasets have been established by capturing real noisy scenes~\cite{plotz2017benchmarking,anaya2018renoir,abdelhamed2018high}, which promotes research into real-image denoising. N3Net~\cite{Ploetz2018NNN} demonstrated the significance on real noisy dataset. CBDNet~\cite{guo2019toward} trained two subnets to sequentially estimate noise and perform non-blind denoising. PD~\cite{zhou2019awgn} applied the pixel-shuffle downsampling strategy to approximate the real noise to AWGN, which can adapt the trained model to real noises. RIDNet~\cite{anwar2019real} proposed a one-stage denoising network with feature attention for real image denoising. However, these methods lack adaptability to image content and result in oversmoothing artifacts.

\section{Framework}

\begin{figure*}[t]
\begin{center}
\includegraphics[width=0.95\linewidth]{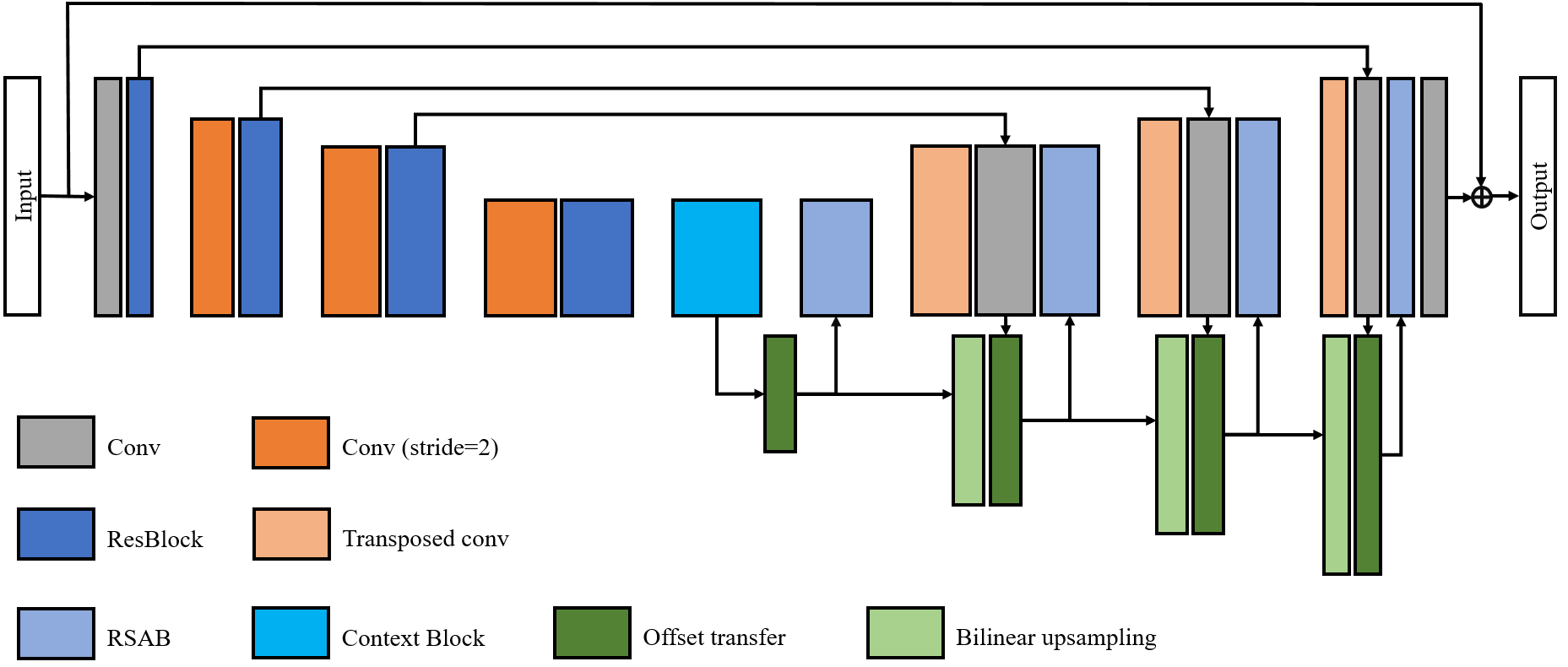}
\end{center}
\setlength{\abovecaptionskip}{0.cm}
  \caption{The framework of our proposed spatial-adaptive denoising network.}
\label{fig_framework}
\end{figure*}

The architecture of our proposed spatial-adaptive denoising network (SADNet) is shown in Fig. \ref{fig_framework}. Let $x$ denotes a noisy input image and $\hat{y}$ denotes the corresponding output denoised image. Then our model can be described as follows:
\begin{equation}
    \hat{y}={\rm SADNet}(x).
\end{equation}
We use one convolutional layer to extract the initial features from the noisy input; then those features are input into a multiscale encoder-decoder architecture. In the encoder component, we use ResBlocks~\cite{he2016deep} to extract features of different scales. However, unlike the original ResBlock, we remove the batch normalization and use leaky ReLU~\cite{maas2013rectifier} as the activation function. To avoid damaging the image structures, we limit the number of downsampling operations and implement a context block to further enlarge the receptive field and capture multiscale information. Then, in the decoder component, we design residual spatial-adaptive blocks (RSABs) to sample and weight the related features to remove noise and reconstruct the textures. In addition, we estimate the offsets and transfer them from coarse to fine, which is beneficial for obtaining more accurate feature locations. Finally the reconstructed features are fed to the last convolutional layer to restore the denoised image. By using the long residual connection, our network learns only the noise component.

In addition to the network architecture, the loss function is crucial to the performance. Several loss functions, such as $L_2$~\cite{zhang2017beyond,zhang2018ffdnet,zhang2019residual}, $L_1$~\cite{anwar2019real}, perceptual loss~\cite{Jiao2017FormResNet}, and asymmetric loss~\cite{guo2019toward}, have been used in denoising tasks. In general, $L_1$ and $L_2$ are the two losses used most commonly in previous works. The $L_2$ loss has good confidence for Gaussian noise, whereas the $L_1$ loss has better tolerance for outliers. 
In our experiment, we use the $L_2$ loss for training on synthetic image datasets and the $L_1$ loss for training on real-image noise datasets.

The following subsections focus on the RSAB and context block to provide more detailed explanations.

\subsection{Residual spatial-adaptive block}


\begin{figure}[t]
\begin{center}
\includegraphics[width=0.7\linewidth]{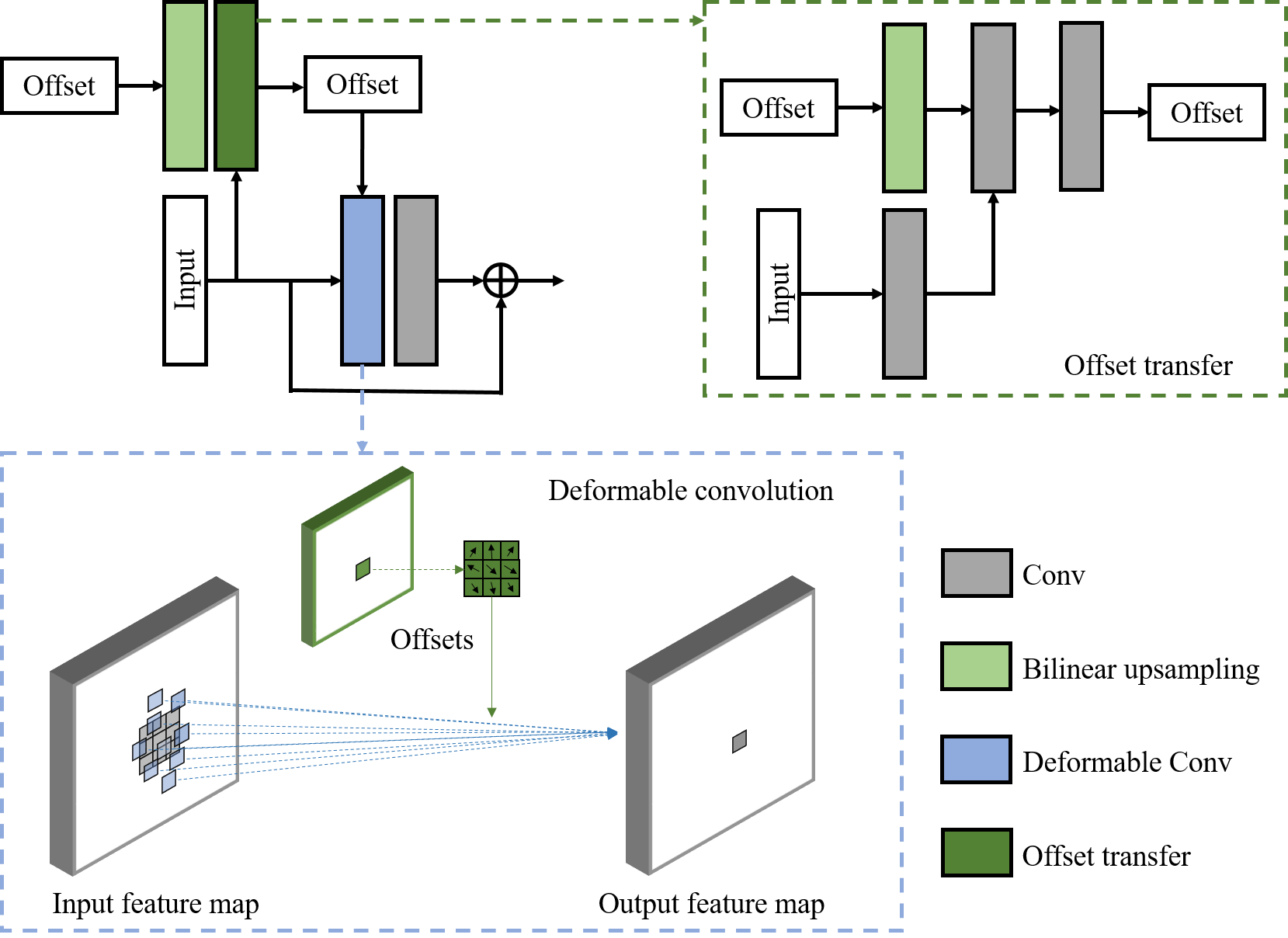}
\end{center}
\setlength{\abovecaptionskip}{0.cm}
   \caption{\textbf{The architecture of the residual spatial-adaptive block (RSAB).} The offset transfer component is shown in the green dashed box. The deformable convolution architecture is shown in the blue dashed box.}
\label{fig_RSAB}
\end{figure}

In this section, we first introduce the deformable convolution~\cite{dai2017deformable,zhu2019deformable} and then propose our RSAB in detail.

Let $x(p)$ denote the features at location $p$ from the input feature map $x$. Then, for a traditional convolution operation, the corresponding output features $y(p)$ can be obtained by
\begin{equation}
    y(p)=\sum_{p_i\in N(p)} w_i \cdot x(p_i),
\end{equation}
where $N(p)$ denotes the neighborhood of location $p$, whose size is equal to the size of the convolutional kernel. $w_i$ denotes the weight of location $p$ in the convolutional kernel, and $p_i$ denotes the location in $N(p)$. The traditional convolution operation strictly takes the feature of the fixed location around $p$ when calculating the output feature. Thus, some unwanted or unrelated features can interfere with the output calculation. For example, when the current location is near the edge, the distinct features located outside the object are introduced for weighting, which may smooth the edges and destroy the texture. For the denoising task, we would prefer that only the related or similar features are used for noise removal, similar to the self-similarity weighted denoising methods~\cite{buades2005non,dabov2007color,dabov2007image}. 

Therefore, we introduce deformable convolution~\cite{dai2017deformable,zhu2019deformable} to adapt to spatial texture changes. In contrast to traditional convolutional layers, deformable convolution can change the shapes of convolutional kernels. It first learns an offset map for every location and applies the resulting offset map to the feature map, which resamples the corresponding features for weighting. Here, we use modulated deformable convolution~\cite{zhu2019deformable}, which provides another dimension of freedom to adjust its spatial support regions, 
\begin{equation}
   y(p) = \sum_{p_i\in N(p)} w_i \cdot x(p_i+\Delta p_i) \cdot \Delta m_i,
\end{equation}
where $\Delta p_i$ is the learnable offset for location $p_i$, and $\Delta m_i$ is the learnable modulation scalar, which lies in the range $[0,1]$. It reflects the degree of correlation between the sampled features $x(p_i)$ and the features in the current location. Thus, the modulated deformable convolution can modulate the input feature amplitudes to further adjust the spatial support regions. Both $\Delta p$ and $\Delta m$ are obtained from the previous features. 

In each RSAB, we first fuse the extracted features and the reconstructed features from the previous scale as the input. The RSAB is constructed by a modulated deformable convolution followed by a traditional convolution with a short skip connection. Similar to ResBlock, we implement local residual learning to enhance the information flow and improve representation ability of the network. However, unlike ResBlock, we replace the first convolution with modulated deformable convolution and use leaky ReLU as our activation function. Hence, the RSAB can be formulated as
\begin{equation}
    F_{RSAB}(x) = F_{cn}(F_{act}(F_{dcn}(x))) + x,
\end{equation}
where $F_{dcn}$ and $F_{cn}$ denote the modulated deformable convolution and traditional convolution respectively. $F_{act}$ is the activation function (leaky ReLU here). The architecture of RSAB is shown in Fig. \ref{fig_RSAB}.

Furthermore, to better estimate the offsets from coarse to fine, we transfer the last-scale offsets $\Delta p^{s-1}$ and modulation scalars $\Delta m^{s-1}$ to the current scale $s$, and then use both $\{\Delta p^{s-1}, \Delta m^{s-1}\}$ and the input features $x^s$ to estimate $\{\Delta p^s, \Delta m^s\}$. Given the small-scale offsets as the initial reference, the related features can be located more accurately on the large scale. The offset transfer can be formulated as follows:
\begin{equation}
    \{\Delta p^s, \Delta m^s\} = F_{offset}(x, F_{up}(\{\Delta p^{s-1}, \Delta m^{s-1}\})),
\end{equation}
where $F_{offset}$ and $F_{up}$ denote the offset transfer and upsampling functions, separately, as shown in Fig. \ref{fig_RSAB}. The offset transfer function involves several convolutions, and it extracts features from input and fuses them with the previous offsets to estimate the offsets in the current scale. The upsampling function magnifies both the size and value of the previous offset maps. In our experiment, bilinear interpolation is adopted to upsample the offsets and modulation scalars.

\subsection{Context block}

\begin{figure}[t]
\begin{center}
\includegraphics[width=0.5\linewidth]{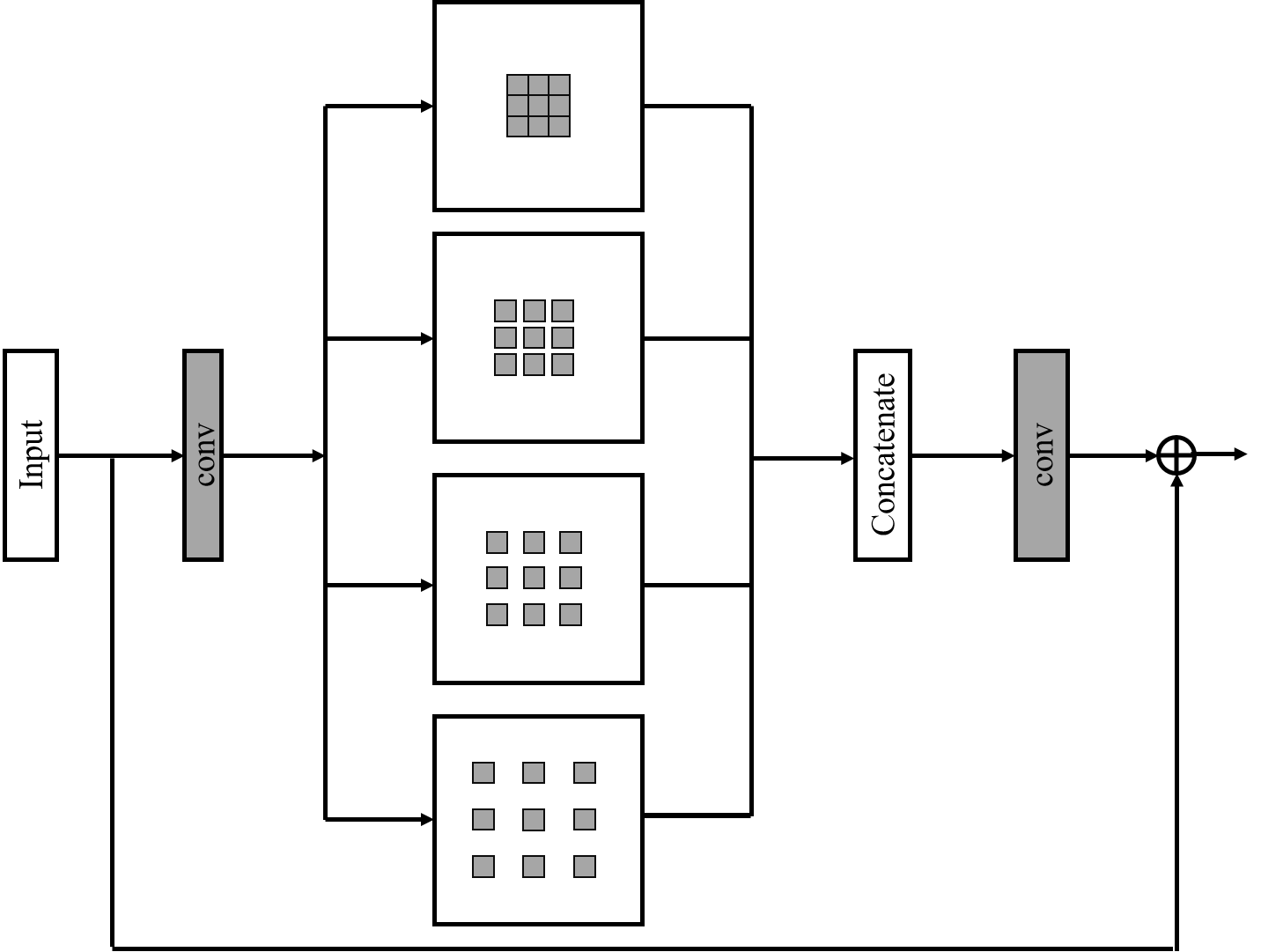}
\end{center}
\setlength{\abovecaptionskip}{0.cm}
   \caption{\textbf{The architecture of the context block.} Instead of downsampling operations, multisize dilated convolutions are implemented to extract different receptive-field features.}
\label{fig_context}
\end{figure}

Multiscale information is important for image denoising tasks; therefore, the downsampling operation is often adopted in networks. However, when the spatial resolution is too small, the image structures are destroyed, and information is lost, which is not conducive to reconstructing the features.

To increase the receptive field and capture multiscale information without further reducing the spatial resolution, we introduce a context block into the minimum scale between the encoder and decoder. Context blocks have been successfully used in image segments~\cite{chen2017rethinking} and deblurring tasks~\cite{zhou2019davanet}. In contrast to spatial pyramid pooling~\cite{he2015spatial}, the context block uses several dilated convolutions with different dilation rates rather than downsampling. 
It can expand the receptive field without increasing the number of parameters or damaging the structures. Then, the features extracted from the different receptive fields are fused to estimate the output (as shown in Fig. \ref{fig_context}). It is beneficial to estimate offsets from a larger receptive field. 

In our experiment, we remove the batch normalization layer and only use four dilation rates which are set to 1, 2, 3, and 4. To further simplify the operation and reduce the running time, we first use a $1\times 1$ convolution to compress the feature channels. The compression ratio is set to 4 in our experiments. In the fusion setup, we use a $1\times 1$ convolution to output the fusion features whose channels are equal to the original input features. Similarly, a local skip connection between the input and output features is applied to prevent information blocking.

\subsection{Implementation}

In the proposed model, we use four scales for the encoder-decoder architecture, and the number of channels for each scale is set to 32, 64, 128, and 256. The kernel size of the first and last convolutional layers is set to $1\times 1$, and the final output is set to 1 or 3 channels depending on the input. Moreover, we use $2\times 2$ filters for up/down-convolutional layers, and all the other convolutional layers have a kernel size of $3\times 3$.
\section{Experiments}

In this section, we demonstrate the effectiveness of our model on both synthetic datasets and real noisy datasets. We adopt DIV2K~\cite{martin2001database} which contains 800 images with 2K resolution, and add different levels of noise to synthetic noise datasets. For real noisy images, we use the SIDD~\cite{abdelhamed2018high}, RENOIR~\cite{anaya2018renoir} and Poly~\cite{xu2018real} datasets. We randomly rotate and flip the images horizontally and vertically for data augmentation. In each training batch, we use 16 patches with size of $128 \times 128$ as inputs. We train our model using the ADAM~\cite{kingma2014adam} optimizer with $\beta_1=0.9$, $\beta_2=0.999$, and $\epsilon=10^{-8}$. The initial learning rate is set to $10^{-4}$ and then halved after $3\times 10^5$ iterations. Our model is implemented in the PyTorch framework~\cite{paszke2019pytorch} with an Nvidia GeForce RTX 1080Ti. In addition, we employ PSNR and SSIM~\cite{wang2004image} to evaluate the results.

\subsection{Ablation study}

We perform ablation study on the Kodak24 dataset with a noise sigma of 50. The results are shown in Table \ref{table_ablation}.

\begin{table}
\setlength{\tabcolsep}{4pt}
\begin{center}
\caption{Ablation study of different components. PSNR values are based on Kodak24 ($\sigma=50$)}
\label{table_ablation}
\begin{tabular}{lcccccc}
\hline\noalign{\smallskip}
RSAB & $\times$ & \checkmark & $\times$ & \checkmark & \checkmark & \checkmark\\
offset transfer & $\times$ & $\times$ & $\times$ & $\times$ & \checkmark & \checkmark\\
Context block & $\times$ & $\times$ & \checkmark & \checkmark & $\times$ & \checkmark \\
\hline\noalign{\smallskip}
PSNR & 29.05 & 29.55 & 29.12 & 29.60 & 29.59 & 29.64\\
\hline
\end{tabular}
\end{center}
\end{table}

\textbf{Ablation on RSAB} RSAB is the crucial block in our network. Without it, the network will lose its ability to adapt to image content. When we replace RSAB with an original ResBlock, the performance decreases substantially, which demonstrates its effect.

\textbf{Ablation on the context block} The context block complements the downsampling operations to capture larger field information. We can observe that the performance improves when the context block is introduced.

\textbf{Ablation on the offset transfer} We remove the offset transfer from coarse to fine and use only the features on the current scale to estimate the offsets for RSAB. This comparison validates the effectiveness of offset transfer.

\subsection{Analyses of the spatial adaptability}
\begin{figure}[t]
\begin{center}
\includegraphics[width=1\linewidth]{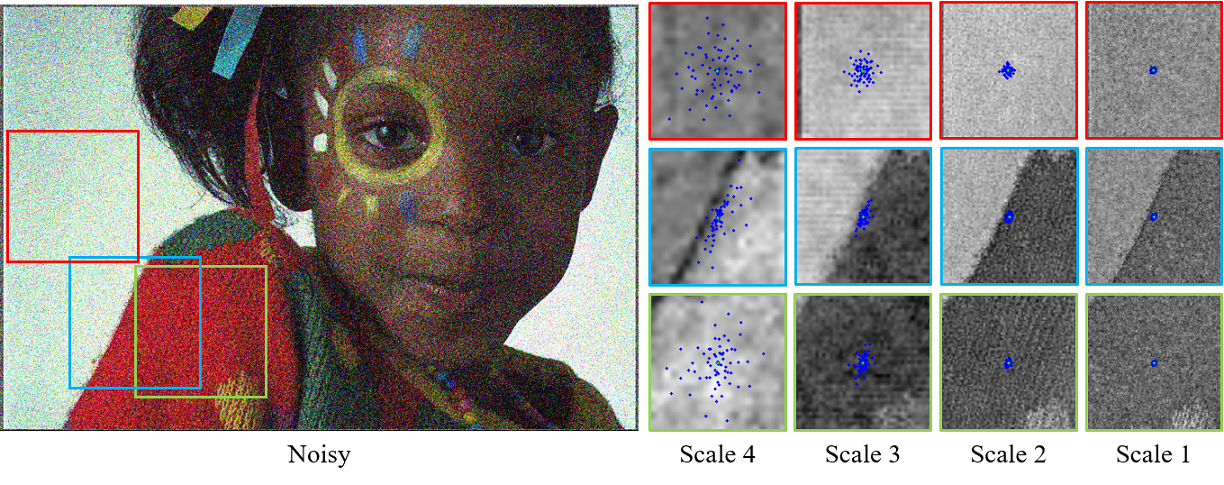}
\end{center}
\setlength{\abovecaptionskip}{0.cm}
   \caption{\textbf{Visualization of the learned kernels.} The scales from 4 to 1 are in order from coarse to fine.}
\label{fig_kernel}
\end{figure}

As discussed above, our network introduces the adaptability to spatial textures and edges. The RSABs can extract related features by change the sampling locations based on the image content. We visualize the learned kernel locations of the RSABs in Fig.~\ref{fig_kernel}. The visualization results show that in the smooth regions or the homogeneous textured regions, the convolution kernels are approximately uniformly distributed, while in the regions close to the edge, the shapes of the convolution kernels extend along the edge. Most of sampling points fall on the similar texture regions inside the object, which demonstrates that our network has indeed learned spatial adaptability. Moreover, as shown in Fig.~\ref{fig_kernel}, the RSAB can extract features from a larger receptive field at the coarse scale, while at the fine scale, the sampled features are located in the neighborhood of the current point. The multiscale structure enables the network to obtain the information of different receptive fields for image reconstruction.    

\subsection{Comparisons}

In this subsection, we compare our algorithm with the state-of-the-art denoising methods. For a fair comparison, all the compared methods employ the default settings provided by the corresponding authors. We first make a comparison on the synthetic noise datasets, since many methods provide only Gaussian noise removal results. Then, we report the denoising results on the real noisy datasets using the state-of-the-art real noise removal methods.

\subsubsection{Synthetic noisy images}

In the comparisons of synthetic noisy images, we use BSD68 and Kodak24 as our test datasets. These datasets include both color and grayscale images for testing. We add AWGN at different noise levels to the clean images. We choose BM3D~\cite{dabov2007image} and  CBM3D~\cite{dabov2007color} as representatives of the classical traditional methods as well as some CNN-based methods, including DnCNN~\cite{zhang2017beyond}, MemNet~\cite{tai2017memnet}, FFDNet~\cite{zhang2018ffdnet}, RNAN~\cite{zhang2019residual}, and  RIDNet~\cite{anwar2019real}, for the comparisons.

Tables \ref{table_gray} shows the average results of PSNR on grayscale images with three different noise levels. Our SADNet  achieves the highest values on most of the datasets and tested noise levels. Note that although RNAN can achieve comparable evaluations to our method on partial low noise levels, it requires more parameters and a larger computational overhead. Next, Table \ref{table_color} reports the quantitative results on color images. We replace the input and output channels from one to three as the other methods. Our SADNet outperforms the state-of-the-art methods on all the datasets with all tested noise levels. In addition, we can observe that our method shows more improvement at higher noise levels, which demonstrates its effectiveness for heavy noise removal. 

\begin{table*}
\setlength{\tabcolsep}{2pt}
\begin{center}
\caption{Average PSNR(dB) results on synthetic \textbf{grayscale} noisy images}
\label{table_gray}
\begin{tabular}{lcccccccc}
\hline
Dataset & $\sigma$ & BM3D & DnCNN & MemNet & FFDNet & RNAN & RIDNet & SADNet (ours) \\
\hline\hline
\multirow{2}{*}{BSD68} & 30 & 27.76 & 28.36 & 28.43 & 28.39 & 28.61 & 28.54 & \textbf{28.61}\\
~ & 50 & 25.62 & 26.23 & 26.35 & 26.29 & 26.48 & 26.40 & \textbf{26.51}\\
~ & 70 & 24.44 & 24.90 & 25.09 & 25.04 & 25.18 & 25.12 & \textbf{25.24}\\
\hline
\multirow{2}{*}{Kodak24} & 30 & 29.13 & 29.62 & 29.72 & 29.70 & \textbf{30.04} & 29.90 & 30.00\\
~ & 50 & 26.99 & 27.51 & 27.68 & 27.63 & 27.93 & 27.79 & \textbf{27.96}\\
~ & 70 & 25.73 & 26.08 & 26.42 & 26.34 & 26.60 & 26.51 & \textbf{26.72}\\
\hline
\end{tabular}
\end{center}
\end{table*}

\begin{table*}
\setlength{\tabcolsep}{2pt}
\begin{center}
\caption{Average PSNR(dB) results on synthetic \textbf{color} noisy images}
\label{table_color}
\begin{tabular}{lcccccccc}
\hline
Dataset & $\sigma$ & CBM3D & DnCNN & MemNet & FFDNet & RNAN & RIDNet & SADNet (ours) \\
\hline\hline
\multirow{2}{*}{BSD68} & 30 & 29.73 & 30.40 & 28.39 & 30.31 & 30.63 & 30.47 & \textbf{30.64}\\
~ & 50 & 27.38 & 28.01 & 26.33 & 27.96 & 28.27 & 28.12 & \textbf{28.32}\\
~ & 70 & 26.00 & 26.56 & 25.08 & 26.53 & 26.83 & 26.69 & \textbf{26.93}\\
\hline
\multirow{2}{*}{Kodak24} & 30 & 30.89 & 31.39 & 29.67 & 31.39 & 31.86 & 31.64 & \textbf{31.86}\\
~ & 50 & 28.63 & 29.16 & 27.65 & 29.10 & 29.58 & 29.25 & \textbf{29.64}\\
~ & 70 & 27.27 & 27.64 & 26.40 & 27.68 & 28.16 & 27.94 & \textbf{28.28}\\
\hline
\end{tabular}
\end{center}
\end{table*}

\begin{figure*}
\begin{center}
\includegraphics[width=\linewidth]{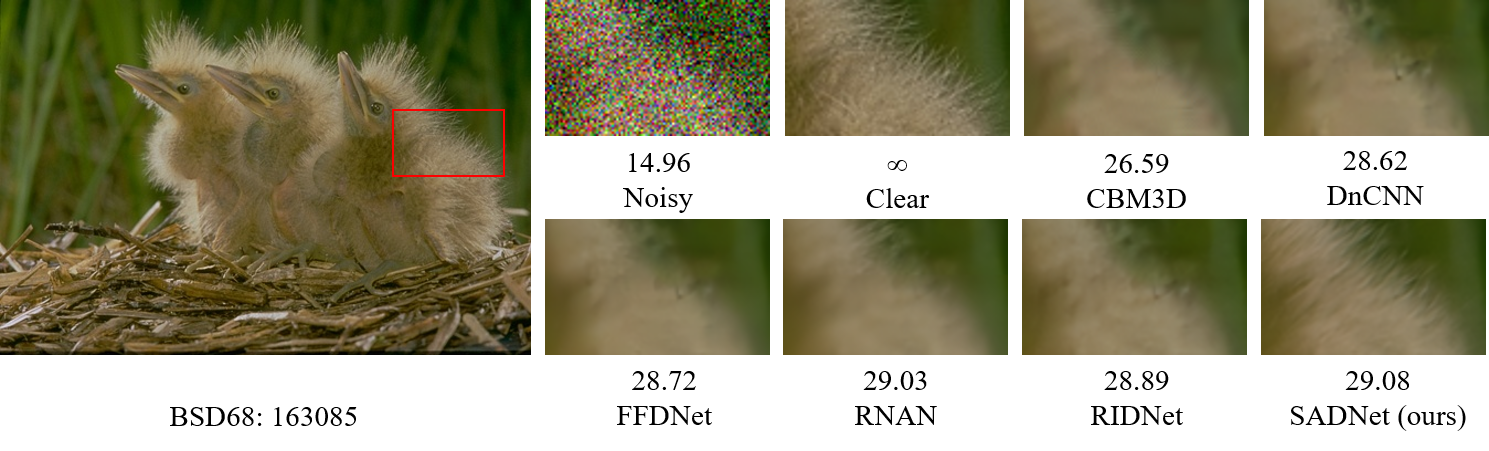}
\end{center}
\setlength{\abovecaptionskip}{0.cm}
   \caption{Synthetic image denoising results on BSD68 with noise level $\sigma=50$.}
\label{fig_bsd68}
\end{figure*}

\begin{figure*}
\begin{center}
\includegraphics[width=\linewidth]{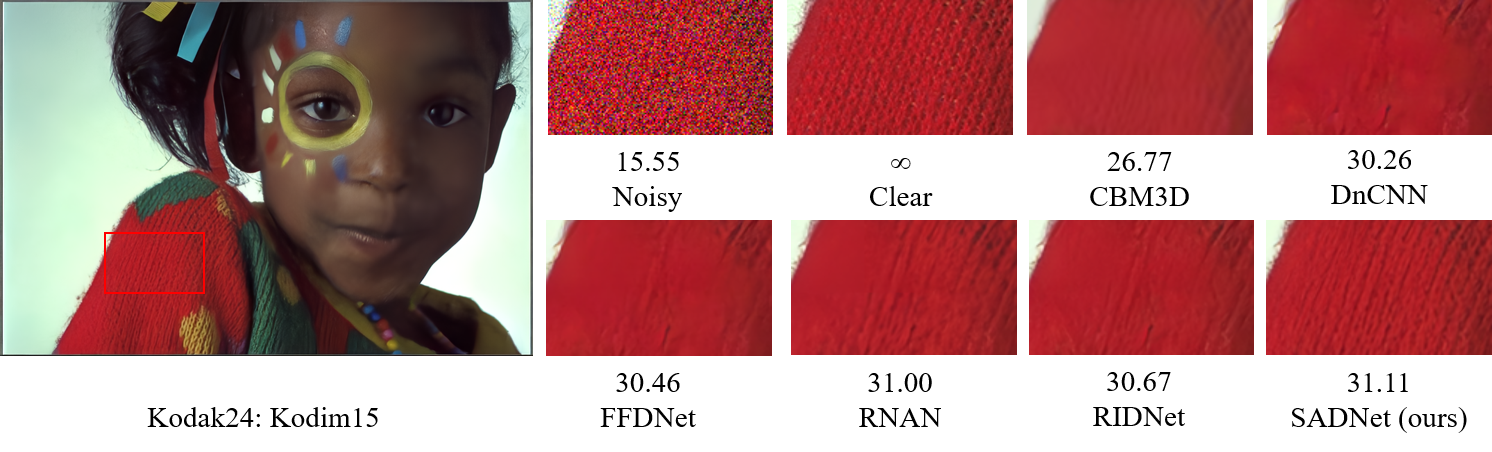}
\end{center}
\setlength{\abovecaptionskip}{0.cm}
   \caption{Synthetic image denoising results on Kodak24 with noise level $\sigma=50$.}
\label{fig_kodak}
\end{figure*}

The visual comparisons are shown in Fig. \ref{fig_bsd68} and Fig. \ref{fig_kodak}. We present some challenging examples from BSD68 and Kodak24. In particular, the birds' feathers and the clothing textures are difficult to separate from heavy noise. The compared methods tend to remove the details along with the noise, resulting in oversmoothing artifacts. Many of the textured areas are heavily smeared in the denoising results. Due to its adaptivity to the image content, our method can restore the vivid textures from noisy images without introducing other artifacts.

\subsubsection{Real noisy images}

To conduct comparisons on real noisy images, we choose DND~\cite{plotz2017benchmarking}, SIDD~\cite{abdelhamed2018high} and Nam~\cite{nam2016holistic} as test datasets. \textbf{DND} contains 50 real noisy images and their corresponding clear images. One thousand patches with a size of $512\times 512$ are extracted from the dataset by the providers for testing and comparison purposes. Since the ground truth images are not publicly available, we can  obtain only the PSNR/SSIM results though the online submission system introduced by~\cite{plotz2017benchmarking}. The validation dataset of \textbf{SIDD} is introduced for our evaluation, which contains 1280 $256\times 256$ noisy-clean image pairs. \textbf{Nam} includes 15 large image pairs with JPEG compression for 11 scenes. We cropped the images into $512\times 512$ patches and selected 25 patches picked by CBDNet~\cite{guo2019toward} for testing. 

We train our model on the SIDD medium dataset and RENOIR for evaluation on the DND and SIDD validation datasets. Then, we finetune our model on the Poly~\cite{xu2018real} for Nam, which improves the performance on the noisy images with JPEG compression.
Furthermore, as comparisons, we choose the state-of-the-art methods whose validity has previously been demonstrated on real noisy images, including CBM3D~\cite{dabov2007color}, DnCNN~\cite{zhang2017beyond}, CBDNet~\cite{guo2019toward}, PD~\cite{zhou2019awgn}, and RIDNet~\cite{anwar2019real}.

\begin{figure*}[t]
\begin{center}
\includegraphics[width=\linewidth]{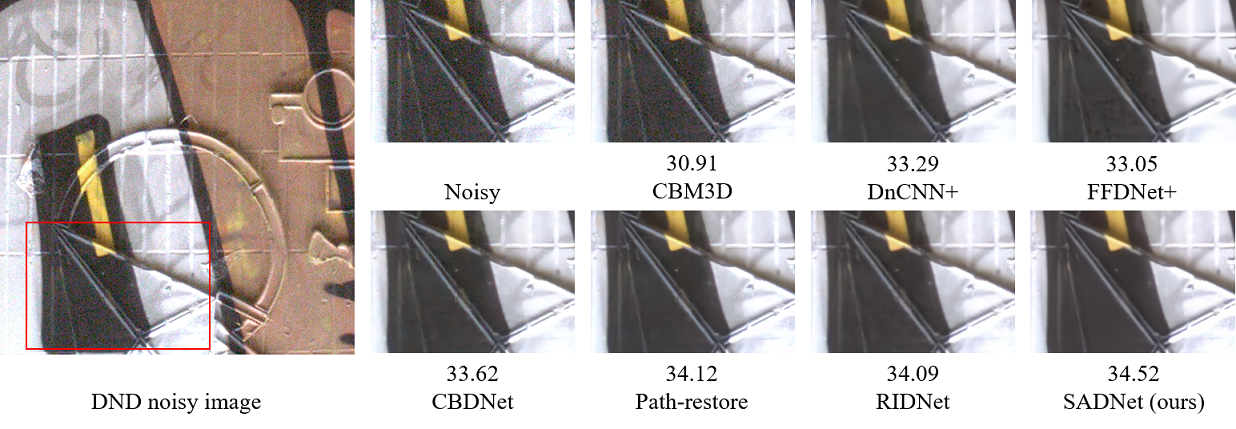}
\end{center}
\setlength{\abovecaptionskip}{0.cm}
   \caption{Real image denoising results from the DnD dataset.}
\label{fig_dnd}
\end{figure*}

\textbf{DND}
The quantitative results are listed in Table \ref{table_DND}, which are obtained from the public DnD benchmark website. FFDNet+ is the improved version of FFDNet with a uniform noise level map manually selected by the providers. CDnCNN-B is the original DnCNN model for blind color denoising. DnCNN+ is finetuned on CDnCNN-B with the results of FFDNet+. SADNet (1248) is the modified version of our SADNet with 1, 2, 4, 8 dilation rates in the context block. Both non-blind and blind denoising methods are included for comparisons. CDnCNN-B cannot effectively generalize to real noisy images. The performances of non-blind denoising methods are limited due to the different distributions between AWGN and real-world noise. In contrast, our SADNet outperforms the state-of-the-art methods with respect to both PSNR and SSIM values. We further perform a visual comparison on denoised images from the DnD dataset, as shown in Fig. \ref{fig_dnd}. The other methods corrode the edges with residual noise, while our method can effectively remove the noise from the smooth region and maintain clear edges.     

\begin{table}
\setlength{\tabcolsep}{6pt}
\begin{center}
\caption{Quantitative results on DnD sRGB images}
\label{table_DND}
\begin{tabular}{lccc}
\hline
Method & Blind/Non-blind & PSNR & SSIM \\
\hline\hline
CDnCNN-B & Blind & 32.43 & 0.7900\\
TNRD & Non-blind & 33.65 & 0.8306\\
BM3D & Non-blind & 34.51 & 0.8507\\
WNNM & Non-blind & 34.67 & 0.8646\\
MCWNNM & Non-blind & 37.38 & 0.9294\\
FFDNet+ & Non-blind & 37.61 & 0.9415\\
DnCNN+ & Non-blind & 37.90 & 0.9430\\
CBDNet & Blind & 38.06 & 0.9421\\
N3Net & Blind & 38.32 & 0.9384\\
PD & Blind & 38.40 & 0.9452\\
Path-Restore & Blind & 39.00 & 0.9542\\
RIDNet & Blind & 39.26 & 0.9528\\
\hline
SADNet (1248) & Blind & 39.37 & \textbf{0.9544}\\
SADNet (ours) & Blind & \textbf{39.59} & 0.9523\\
\hline
\end{tabular}
\end{center}
\end{table}

\begin{table}
\setlength{\tabcolsep}{2pt}
\begin{center}
\caption{Quantitative results on SIDD sRGB validation dataset}
\label{table_SIDD}
\begin{tabular}{lcccccc}
\hline
Method & CBM3D & CDnCNN-B & CBDNet & PD & RIDNET & SADNet (ours)\\
\hline\hline
Blind/Non-blind & Non-blind & Blind & Blind & Blind & Blind & Blind \\
PSNR & 30.88 & 26.21 & 30.78 & 32.94 & 38.71 & \textbf{39.46}\\
\hline
\end{tabular}
\end{center}
\end{table}

\begin{figure}
\begin{center}
\includegraphics[width=0.95\linewidth]{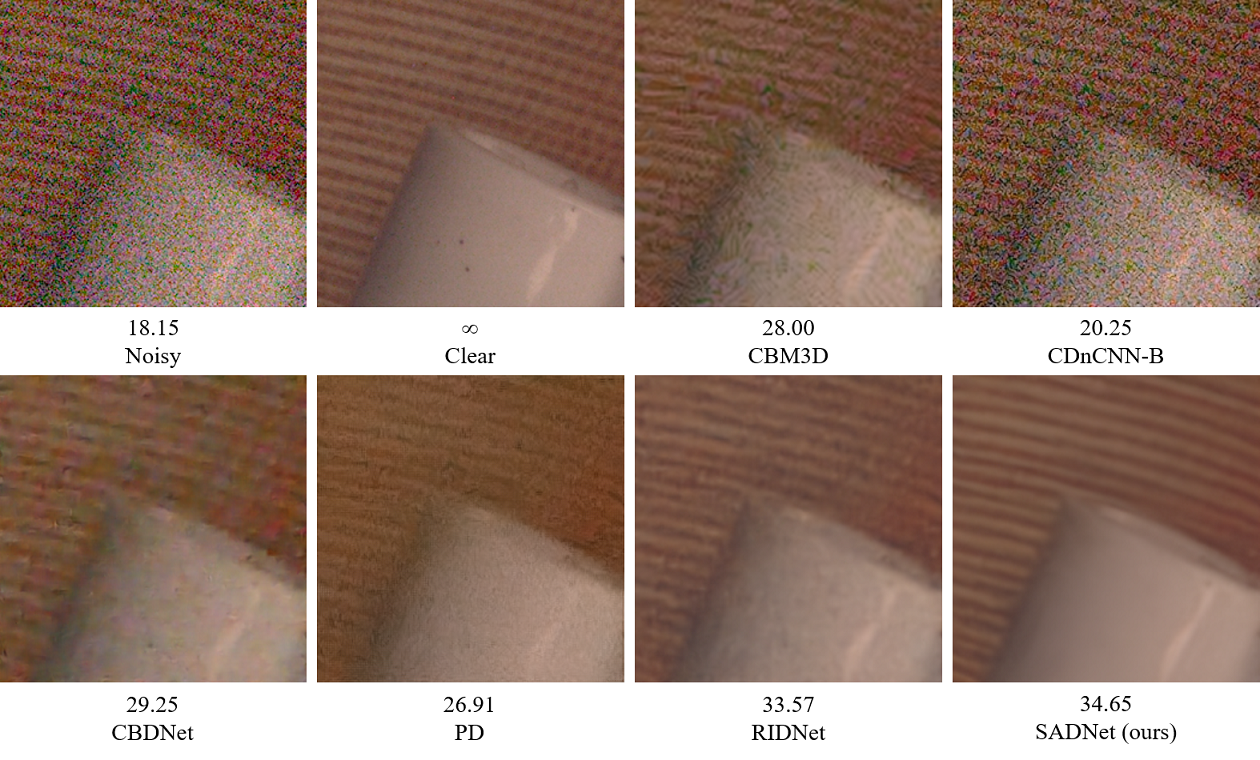}
\end{center}
\setlength{\abovecaptionskip}{0.cm}
   \caption{A real image denoising example from the SIDD dataset.}
\label{fig_sidd_1}
\end{figure}

\begin{figure}
\begin{center}
\includegraphics[width=0.95\linewidth]{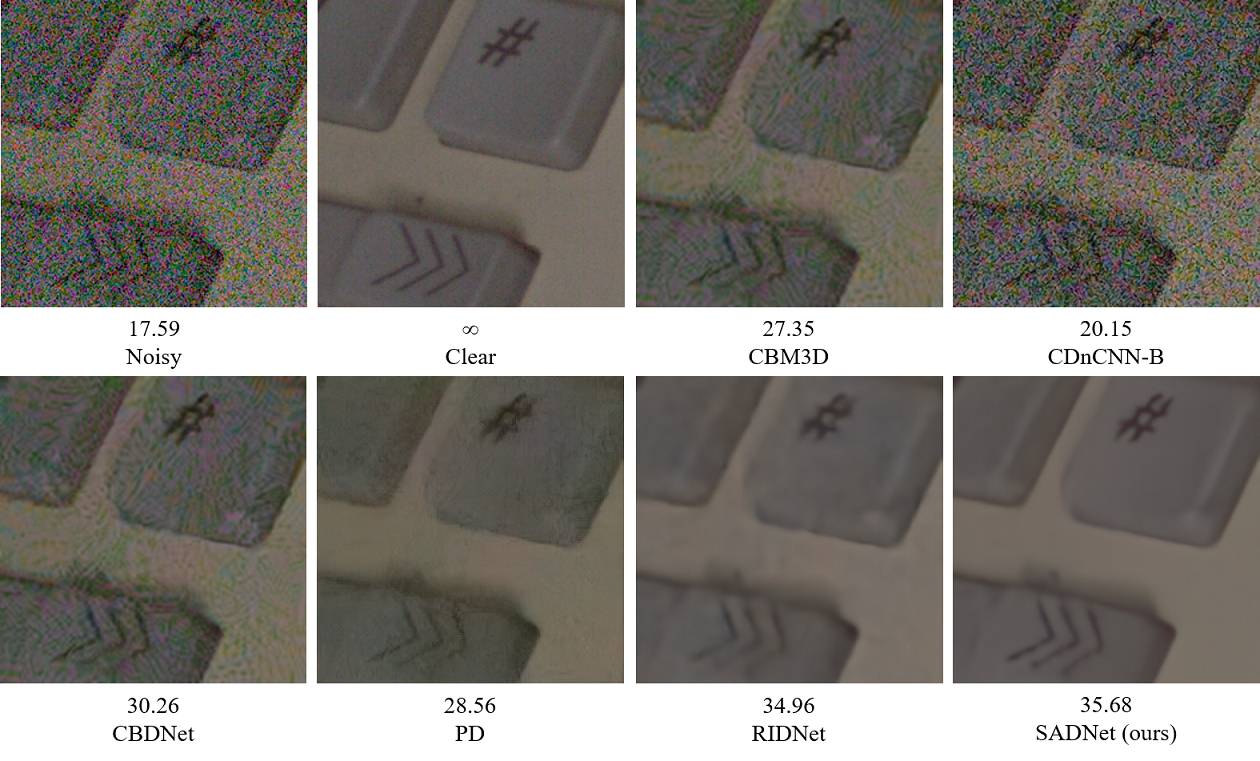}
\end{center}
\setlength{\abovecaptionskip}{0.cm}
   \caption{Another real image denoising example from the SIDD dataset.}
\label{fig_sidd_2}
\end{figure}

\textbf{SIDD}
The images in the SIDD dataset are captured by smartphones, and some noisy images have high noise levels. We employ 1,280 validation images for quantitative comparisons as listed in Table \ref{table_SIDD}. The results demonstrates that our method achieves significant improvements over the other tested methods. For visual comparisons, we choose two challenging examples from the denoised results. The first scene has rich textures, while the second scene has prominent structures. As shown in Fig. \ref{fig_sidd_1} and Fig. \ref{fig_sidd_2}, CDnCNN-B and CBDNet fail at noise removal. CBM3D results in pseudo artifacts, and PD and RIDNet destroy the textures. In contrast, our network recovers textures and structures that are closer to the ground truth.

\begin{figure}
\begin{center}
\includegraphics[width=\linewidth]{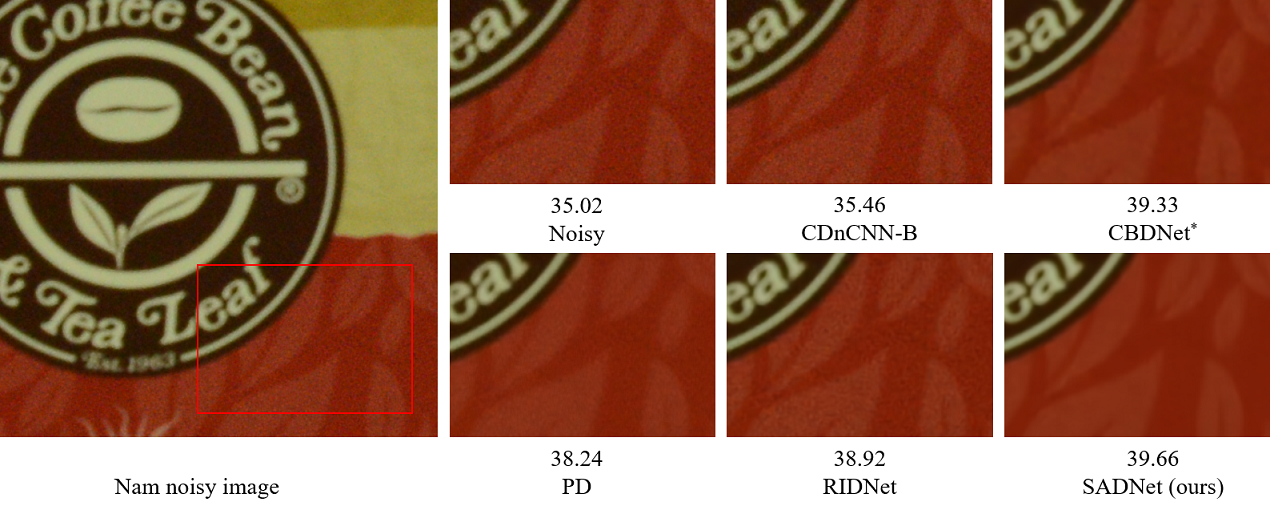}
\end{center}
\setlength{\abovecaptionskip}{0.cm}
   \caption{Real image denoising results from the Nam dataset with JPEG compression.}
\label{fig_nam}
\end{figure}

\textbf{Nam}
The JPEG compression makes the noise more stubborn on the Nam dataset. For a fair comparison, we use the patches chosen by CBDNet~\cite{guo2019toward} for evaluation. Furthermore, CBDNet*~\cite{guo2019toward} is introduced for comparison, which was retrained on JPEG compressed datasets by its providers. We report the average PSNR and SSIM values for Nam in Table \ref{table_Nam}. With respect to PSNR, Our SADNet achieves 1.88, 1.83 and 1.61 dB gains over RIDNet, PD, and CBDNet*. Similarly, our SSIM values exceed those of all the other methods in the comparison. In the visual comparison shown in Fig. \ref{fig_nam}, our method again obtains the best result for texture restoration and noise removal.

\begin{table}
\setlength{\tabcolsep}{2pt}
\begin{center}
\caption{Quantitative results on Nam dataset with JPEG compression}
\label{table_Nam}
\begin{tabular}{lcccccc}
\hline
Method & CBM3D & CDnCNN-B & CBDNet* & PD & RIDNET & SADNet (ours)\\
\hline\hline
Blind/Non-blind & Non-blind & Blind & Blind & Blind & Blind & Blind \\
PSNR & 39.84 & 37.49 & 41.31 & 41.09 & 41.04 & \textbf{42.92}\\
SSIM & 0.9657 & 0.9272 & 0.9784 & 0.9780 & 0.9814 & \textbf{0.9839} \\
\hline
\end{tabular}
\end{center}
\end{table}

\begin{table*}
\setlength{\tabcolsep}{4pt}
\begin{center}
\caption{Parameters and time comparisons on $480\times 320$ color images}
\label{table_time}
\begin{tabular}{lccccc}
\hline
Method & DnCNN & MemNet & RNAN & RIDNet & SADNET (ours)\\
\hline\hline
Parameters & 558k & 2,908k & 8,960k & 1,499k & 4,321k \\
FLOPs & 86.1G & 449.2G & 1163.5G & 230.0G & 50.1G \\
times (ms) & 21.3 & 154.2 & 1072.2 & 84.4 & 26.7 \\
\hline
\end{tabular}
\end{center}
\end{table*}

\subsubsection{Parameters and running times}

To compare the running times, we test different methods when denoising $480\times 320$ color images. Note that the running time may depend on the test platform and code; thus, we also provide the number of floating point operations (FLOPs). All the methods are implemented in PyTorch. As shown in Table~\ref{table_time}, although SADNet has high parameter numbers, its FLOPs are minimal, and its running time is short due to the multiple downsampling operations. Because most of the operations run on smaller-scale feature maps, our model performs faster than many others with fewer parameters.

\section{Conclusion}

In this paper, we propose a spatial-adaptive denoising network for effective noise removal. The network is built by multiscale residual spatial-adaptive blocks, which sample relevant features for weighting based on the content and textures of images. We further introduce a context block to capture multiscale information and implement offset transfer to more accurately estimate the sampling locations. We find that the introduction of spatially adaptive capability can restore richer details in complex scenes under heavy noise. The proposed SADNet achieves state-of-the-art performances on both synthetic and real noisy images and has a moderate running time.   

\subsubsection{Acknowledgments}

This work is partially supported by Science and Technology on Optical Radiation Laboratory (61424080211).

%
%
\bibliographystyle{splncs04}
\bibliography{egbib}

\begin{thebibliography}{10}
\providecommand{\url}[1]{\texttt{#1}}
\providecommand{\urlprefix}{URL }
\providecommand{\doi}[1]{https://doi.org/#1}

\bibitem{abdelhamed2018high}
Abdelhamed, A., Lin, S., Brown, M.S.: A high-quality denoising dataset for
  smartphone cameras. In: Proceedings of the IEEE Conference on Computer Vision
  and Pattern Recognition. pp. 1692--1700 (2018)

\bibitem{aharon2006k}
Aharon, M., Elad, M., Bruckstein, A.: K-svd: An algorithm for designing
  overcomplete dictionaries for sparse representation. IEEE Transactions on
  signal processing  \textbf{54}(11),  4311--4322 (2006)

\bibitem{anaya2018renoir}
Anaya, J., Barbu, A.: Renoir--a dataset for real low-light image noise
  reduction. Journal of Visual Communication and Image Representation
  \textbf{51},  144--154 (2018)

\bibitem{anwar2019real}
Anwar, S., Barnes, N.: Real image denoising with feature attention. arXiv
  preprint arXiv:1904.07396  (2019)

\bibitem{buades2005non}
Buades, A., Coll, B., Morel, J.M.: A non-local algorithm for image denoising.
  In: 2005 IEEE Computer Society Conference on Computer Vision and Pattern
  Recognition (CVPR'05). vol.~2, pp. 60--65. IEEE (2005)

\bibitem{chen2017rethinking}
Chen, L.C., Papandreou, G., Schroff, F., Adam, H.: Rethinking atrous
  convolution for semantic image segmentation. arXiv preprint arXiv:1706.05587
  (2017)

\bibitem{chen2016trainable}
Chen, Y., Pock, T.: Trainable nonlinear reaction diffusion: A flexible
  framework for fast and effective image restoration. IEEE transactions on
  pattern analysis and machine intelligence  \textbf{39}(6),  1256--1272 (2016)

\bibitem{dabov2007color}
Dabov, K., Foi, A., Katkovnik, V., Egiazarian, K.: Color image denoising via
  sparse 3d collaborative filtering with grouping constraint in
  luminance-chrominance space. In: 2007 IEEE International Conference on Image
  Processing. vol.~1, pp. I--313. IEEE (2007)

\bibitem{dabov2007image}
Dabov, K., Foi, A., Katkovnik, V., Egiazarian, K.: Image denoising by sparse
  3-d transform-domain collaborative filtering. IEEE Transactions on image
  processing  \textbf{16}(8),  2080--2095 (2007)

\bibitem{dai2017deformable}
Dai, J., Qi, H., Xiong, Y., Li, Y., Zhang, G., Hu, H., Wei, Y.: Deformable
  convolutional networks. In: Proceedings of the IEEE international conference
  on computer vision. pp. 764--773 (2017)

\bibitem{gu2014weighted}
Gu, S., Zhang, L., Zuo, W., Feng, X.: Weighted nuclear norm minimization with
  application to image denoising. In: Proceedings of the IEEE conference on
  computer vision and pattern recognition. pp. 2862--2869 (2014)

\bibitem{guo2019toward}
Guo, S., Yan, Z., Zhang, K., Zuo, W., Zhang, L.: Toward convolutional blind
  denoising of real photographs. In: Proceedings of the IEEE Conference on
  Computer Vision and Pattern Recognition. pp. 1712--1722 (2019)

\bibitem{he2015spatial}
He, K., Zhang, X., Ren, S., Sun, J.: Spatial pyramid pooling in deep
  convolutional networks for visual recognition. IEEE transactions on pattern
  analysis and machine intelligence  \textbf{37}(9),  1904--1916 (2015)

\bibitem{he2016deep}
He, K., Zhang, X., Ren, S., Sun, J.: Deep residual learning for image
  recognition. In: Proceedings of the IEEE conference on computer vision and
  pattern recognition. pp. 770--778 (2016)

\bibitem{Jiao2017FormResNet}
Jiao, J., Tu, W.C., He, S., Lau, R.W.H.: Formresnet: Formatted residual
  learning for image restoration. In: 2017 IEEE Conference on Computer Vision
  and Pattern Recognition Workshops (CVPRW) (2017)

\bibitem{kingma2014adam}
Kingma, D.P., Ba, J.: Adam: A method for stochastic optimization. arXiv
  preprint arXiv:1412.6980  (2014)

\bibitem{lefkimmiatis2017non}
Lefkimmiatis, S.: Non-local color image denoising with convolutional neural
  networks. In: Proceedings of the IEEE Conference on Computer Vision and
  Pattern Recognition. pp. 3587--3596 (2017)

\bibitem{liu2018non}
Liu, D., Wen, B., Fan, Y., Loy, C.C., Huang, T.S.: Non-local recurrent network
  for image restoration. In: Advances in Neural Information Processing Systems.
  pp. 1673--1682 (2018)

\bibitem{maas2013rectifier}
Maas, A.L., Hannun, A.Y., Ng, A.Y.: Rectifier nonlinearities improve neural
  network acoustic models. In: Proc. icml. vol.~30, p.~3 (2013)

\bibitem{mairal2009non}
Mairal, J., Bach, F.R., Ponce, J., Sapiro, G., Zisserman, A.: Non-local sparse
  models for image restoration. In: ICCV. vol.~29, pp. 54--62. Citeseer (2009)

\bibitem{martin2001database}
Martin, D., Fowlkes, C., Tal, D., Malik, J., et~al.: A database of human
  segmented natural images and its application to evaluating segmentation
  algorithms and measuring ecological statistics. Iccv Vancouver: (2001)

\bibitem{nam2016holistic}
Nam, S., Hwang, Y., Matsushita, Y., Joo~Kim, S.: A holistic approach to
  cross-channel image noise modeling and its application to image denoising.
  In: Proceedings of the IEEE Conference on Computer Vision and Pattern
  Recognition. pp. 1683--1691 (2016)

\bibitem{osher2005iterative}
Osher, S., Burger, M., Goldfarb, D., Xu, J., Yin, W.: An iterative
  regularization method for total variation-based image restoration. Multiscale
  Modeling \& Simulation  \textbf{4}(2),  460--489 (2005)

\bibitem{paszke2019pytorch}
Paszke, A., Gross, S., Massa, F., Lerer, A., Bradbury, J., Chanan, G., Killeen,
  T., Lin, Z., Gimelshein, N., Antiga, L., et~al.: Pytorch: An imperative
  style, high-performance deep learning library. In: Advances in Neural
  Information Processing Systems. pp. 8024--8035 (2019)

\bibitem{plotz2017benchmarking}
Plotz, T., Roth, S.: Benchmarking denoising algorithms with real photographs.
  In: Proceedings of the IEEE Conference on Computer Vision and Pattern
  Recognition. pp. 1586--1595 (2017)

\bibitem{Ploetz2018NNN}
Pl\"otz, T., Roth, S.: Neural nearest neighbors networks. In: Advances in
  Neural Information Processing Systems (NeurIPS) (2018)

\bibitem{tai2017memnet}
Tai, Y., Yang, J., Liu, X., Xu, C.: Memnet: A persistent memory network for
  image restoration. In: Proceedings of the IEEE international conference on
  computer vision. pp. 4539--4547 (2017)

\bibitem{wang2018non}
Wang, X., Girshick, R., Gupta, A., He, K.: Non-local neural networks. In:
  Proceedings of the IEEE Conference on Computer Vision and Pattern
  Recognition. pp. 7794--7803 (2018)

\bibitem{wang2004image}
Wang, Z., Bovik, A.C., Sheikh, H.R., Simoncelli, E.P.: Image quality
  assessment: from error visibility to structural similarity. IEEE transactions
  on image processing  \textbf{13}(4),  600--612 (2004)

\bibitem{xu2007iterative}
Xu, J., Osher, S.: Iterative regularization and nonlinear inverse scale space
  applied to wavelet-based denoising. IEEE Transactions on Image Processing
  \textbf{16}(2),  534--544 (2007)

\bibitem{xu2018real}
Xu, J., Li, H., Liang, Z., Zhang, D., Zhang, L.: Real-world noisy image
  denoising: A new benchmark. arXiv preprint arXiv:1804.02603  (2018)

\bibitem{xu2018external}
Xu, J., Zhang, L., Zhang, D.: External prior guided internal prior learning for
  real-world noisy image denoising. IEEE Transactions on Image Processing
  \textbf{27}(6),  2996--3010 (2018)

\bibitem{xu2018trilateral}
Xu, J., Zhang, L., Zhang, D.: A trilateral weighted sparse coding scheme for
  real-world image denoising. In: Proceedings of the European Conference on
  Computer Vision (ECCV). pp. 20--36 (2018)

\bibitem{xu2017multi}
Xu, J., Zhang, L., Zhang, D., Feng, X.: Multi-channel weighted nuclear norm
  minimization for real color image denoising. In: Proceedings of the IEEE
  International Conference on Computer Vision. pp. 1096--1104 (2017)

\bibitem{zhang2017beyond}
Zhang, K., Zuo, W., Chen, Y., Meng, D., Zhang, L.: Beyond a gaussian denoiser:
  Residual learning of deep cnn for image denoising. IEEE Transactions on Image
  Processing  \textbf{26}(7),  3142--3155 (2017)

\bibitem{zhang2018ffdnet}
Zhang, K., Zuo, W., Zhang, L.: Ffdnet: Toward a fast and flexible solution for
  cnn-based image denoising. IEEE Transactions on Image Processing
  \textbf{27}(9),  4608--4622 (2018)

\bibitem{zhang2019residual}
Zhang, Y., Li, K., Li, K., Zhong, B., Fu, Y.: Residual non-local attention
  networks for image restoration. arXiv preprint arXiv:1903.10082  (2019)

\bibitem{zhou2019davanet}
Zhou, S., Zhang, J., Zuo, W., Xie, H., Pan, J., Ren, J.S.: Davanet: Stereo
  deblurring with view aggregation. In: Proceedings of the IEEE Conference on
  Computer Vision and Pattern Recognition. pp. 10996--11005 (2019)

\bibitem{zhou2019awgn}
Zhou, Y., Jiao, J., Huang, H., Wang, Y., Wang, J., Shi, H., Huang, T.: When
  awgn-based denoiser meets real noises. arXiv preprint arXiv:1904.03485
  (2019)

\bibitem{zhu2019deformable}
Zhu, X., Hu, H., Lin, S., Dai, J.: Deformable convnets v2: More deformable,
  better results. In: Proceedings of the IEEE Conference on Computer Vision and
  Pattern Recognition. pp. 9308--9316 (2019)

\bibitem{zoran2011learning}
Zoran, D., Weiss, Y.: From learning models of natural image patches to whole
  image restoration. In: 2011 International Conference on Computer Vision. pp.
  479--486. IEEE (2011)

\end{thebibliography}
\end{document}